# Software Engineering & Systems Design Nature

Kirill A. Sorudeykin,
Computer Engineering Expert

*Kharkov National University of Radio Electronics, Ukraine*
*Kirill.A.Sorudeykin@ieee.org*

> *We can't solve problems by using the same kind of thinking we used when we created them.*
>
> *Albert Einstein*

## 1. Abstract

*The main problems of Software Engineering appear as a result of incompatibilities. For example, the quality of organization of the production process depends on correspondence with existent resources and on a common understanding of project goals by all team members. Software design is another example. Its successfulness rides on the architecture's conformity with a project's concepts. This is a point of great nicety. All elements should create a single space of interaction. And if the laws of such a space are imperfect, missequencing comes and the concept of a software system fails. We must do our best for this not to happen. To that end, having a subtle perception of systems structures is essential. Such knowledge can be based only on a fresh approach to the logical law.*

## 2. A problem of Design

Currently, science endures the necessity of a qualitative leap. This means the natural conversion to more profound principles. By considering certain facts and looking for regularities, we discover superior laws, which we also call simplification. Meanwhile, it is hard to determine the verity of a selected research approach. Logic does not allow one to imagine a solution integrally until all solution components are identified. However, we can assume the departing point that we should have in our investigation. Laws of nature and laws of science should not contradict each other.

Every occupation has similarities with others. This fact demonstrates the universal laws of thought, meaning that everybody thinks using the same principles – imagination and logic. Any complex problem can be solved by mere thinking and considering every syllable of the matter. When we reflect on intelligence, we understand that the individuality of every person's thinking consists of their particular clarity of notion relations. But there are no differences between the ways of operation with notions by different people.

A fundamental aspect is that the process of stating the problem can be described as the *space creation mechanism* of a certain task. This consists of defining the laws of interactions, i.e. logic, which automatically specifies properties of the objects of a space. As a result, we obtain a closed system with a law of causality. There are only several major questions along these lines; all the rest are just details. Within the bounds of this problem we should consider correlations between notions as a whole, not from the position of a certain element of a space. Therefore, any logical rule can be explained as barely an effect of the *limitation of spaces*. In this way, *we can describe,* not only natural phenomena, but also the very essence of *logic*.

## 3. Universality of algorithmization

Making a chain of logical inferences between various concepts is usually characterized as enclosure and inheritance of concepts. But we cannot construct an exhaustive description of an investigated system using such a way of thinking. That's why the main question remains unresolved: what is the interaction, its nature and place among objective laws. Modern logic can describe only a small part of the validity, being based not on the concept of space, but on its external view – the so-called "laws of formal logic", which cannot allow relations between various spaces to be described in a simple way. There is a more convenient and perspective way - to find the missed parts in the current description of logical laws. Through the characterization of problem statement rules, we can connect such areas as designing and thinking within the bounds of a uniform model.

Software architecture design regularities display the laws of nature that can be applied to any system and describe the balance of relations between concepts. So, it is impossible to determine the nature of design only in a certain application area. We cannot obtain a satisfactory answer without having looked at the problem from a wider point of view than the particular sphere of application.

Historically, we associate logic with the unshakeable laws of nature. Logic is considered a science about "correct" thinking. By means of logic, the truth can be reached. But accurate explanations of concepts, such as truth and correctness, do not exist in any source. Without a doubt, this is a discrepancy of a theoretical science. There should be objective laws that operate outside of the observer and are not influenced by the human mind.

## 4. P-Modeling

As an example, let's consider P-Modeling Framework by INTSPEI. This methodology includes such elements as Reverse Semantic Traceability and Speechless Modeling, which are based on the characteristics of interactions between independent spaces, i.e. people's minds. Even if several people work together within a single team, team members do not automatically understand their colleagues' goals or see the team's general aims. According to P-Modeling, if we try to check the working results of one team, being independent of their working process, and then try to reconstruct an initial task from their work (i.e. determining a goal of the project), we will demonstrate an objective degree of clearness of a given project. Such a methodology demonstrates an "investigation" process of one space by another, showing potential problems and helping professionals to adapt to the realities of the system design process, i.e. teach them how to heed a greater number of interactions' details.

## 5. Reflective management

If a certain team works on a collective project, any member should have their own individual task. When he or she works according to the plan, there is no way to avoid interceptions with the interests of other members of the team. Due to these interactions, the exchange of viewpoints takes place. Team goals are unified by the mutual correlation of personal tasks. Then, each person begins to see more and more clearly, what the whole team is trying to obtain as a result of their work, and, consequently, everyone is able to determine the most effective way of problem solving and to work in synergy with all the other team members. *Interaction between members of a team superposes their goal visions and leads to collective self-organization.* It is necessary to provide an agreed upon flow of information during communication between team members in order to activate such a mechanism.

Thus, we come to the notion of *reflective management* that is directed to this end. The meaning of reflective management is to support such an environment that can be conducive to an entire team utilizing a single vision of a concept of a workable software system. *System concept* is an objective view of the system's structure and limitations. System concept is also a systematic result of ideas, views and activities of all members of a team by a special group (for example, software engineering subdivision). Through this, an individual opinion of each team member is invested into a general concept that any person can use for "*viewpoint synchronizing*". The system concept is steadily improved (evolved) by interactions [of team members] within a group. Every instance of time shows the integral and balanced vision of a system by the whole group.

## 6. The concept of Spaces Limitation

The term "System's concept" means the space of the restrictions imposed on the capabilities and architectural features of a developed system. No space is meaningful outside of causality. Being outside the scope of its limits is to lose necessity, i.e. to lose clarity of the structure of the space. It is the mortgage and one of most dangerous "reefs" of system design.

The destination of any object is an interaction with other objects within a certain space. An object should be surrounded by space. This is its natural requirement, which always holds true of itself. During the research process, the mind tries to come into contact with an observable object as closely as possible, creating a single space to exclude ambiguities in their interactions.

We need to limit our point of view to a single position. The mental process is a process of determining object interrelation, i.e. one object examines another. One space converges to make other spaces a continuation of itself, adapting to them. The object "images" space representations of the other objects, and it uses only the necessary properties of an object.

As we mentioned above, each space needs a purpose to force the objects' trend of reaching optimal positions. We have shown that each space is characterized by a degree of unattainability of optimal positions. Any subject of a human's activity can be considered from such a point of view. The goal lies beyond the limits of a space realizing it. Further, such a schema of sources of logic law will be called "the principle of spaces limitation". Using this principle, we can interpret the phenomena of intellect as a usual display of the trend of nature towards regulation and balancing.

The main goal of any mental activity is a problem statement (i.e. task space research). After accomplishing this task, a problem solution becomes clear. The vagueness of the solution making process is the disorientation in the space of a problem. If this happens, we cannot see the objects of a more enveloping space, and the absolutely optimal positions of the objects are unattainable in a less enveloping space.

## 6. Summary

Following the Laws of Design predetermines the successfulness of any technological project. Formalization of such laws is very important for exact science, as well as the application areas, such as Software Engineering or Applied Mathematics. Following these laws also gives the analytical apparatus the capability to resolve the set of contradictions mentioned above and will help to represent fundamental knowledge more effectively.

The methodology of research can be defined by a kind of scientific questions. In the given work, the basic moments generating difficulties of analytical sciences have been considered. The leading role is played by the problem of the lack of scientific generalization because we need to unite the empirical facts and qualitatively define new knowledge from these facts. We have proposed a new concept of spaces limitation to arrange the theory in a more harmonious way.

We can extend our imaginations about the role of interactions and the validity of the known laws of nature by trying to see new notions and correlations. Only permanent mental space motion can result in optimal objects arrangement of it. But, to achieve these goals, we need to know how to discern real essences of things from formalisms and how to "navigate" through the space of our profound knowledge.